\pdfoutput=1
\documentclass[preprint, 12pt]{elsarticle}

\usepackage{graphicx}
\usepackage{amssymb}

\usepackage[margin=2cm]{geometry}

\newcommand{\myquote}[1]{``#1''}

\usepackage{float}
\usepackage{caption}
\usepackage{booktabs}	
\usepackage{longtable}
\usepackage{array}
\usepackage{multirow}
\usepackage{threeparttable}	
\usepackage{url}
\usepackage{hyperref}

\newcolumntype{C}[1]{>{\centering\let\newline\\\arraybackslash\hspace{0pt}}m{#1}}
\usepackage{subfigure}

\journal{Medical Image Analysis}

\begin{document}

\begin{frontmatter}


\title{CEREBRUM: a fast and fully-volumetric Convolutional Encoder-decodeR for weakly-supervised sEgmentation\\of BRain strUctures from out-of-the-scanner MRI}



\author[unibs]{Dennis Bontempi}
\author[unibs]{Sergio Benini}
\author[unibs]{Alberto Signoroni}
\author[glasgow]{Michele Svanera\corref{cor1}\fnref{fn1}}
\ead{Michele.Svanera at glasgow.ac.uk}

\author[glasgow]{Lars Muckli\fnref{fn1}}

\address[unibs]{Department of Information Engineering, University of Brescia, Italy}
\address[glasgow]{Institute of Neuroscience and Psychology, University of Glasgow, UK}

\cortext[cor1]{Corresponding author}
\fntext[fn1]{These authors contributed equally}

\begin{abstract}
Many functional and structural neuroimaging studies call for accurate morphometric segmentation of different brain structures starting from image intensity values of MRI scans. Current automatic (multi-) atlas-based segmentation strategies often lack accuracy on difficult-to-segment brain structures and, since these methods rely on atlas-to-scan alignment, they may take long processing times. Alternatively, recent methods deploying solutions based on Convolutional Neural Networks (CNNs) are enabling the direct analysis of out-of-the-scanner data. However, current CNN-based solutions partition the test volume into 2D or 3D patches, which are processed independently. This process entails a loss of global contextual information, thereby negatively impacting the segmentation accuracy. In this work, we design and test an optimised end-to-end CNN architecture that makes the exploitation of global spatial information computationally tractable, allowing to process a whole MRI volume at once. We adopt a weakly supervised learning strategy by exploiting a large dataset composed of 947 out-of-the-scanner (3 Tesla T1-weighted 1mm isotropic MP-RAGE 3D sequences) MR Images. The resulting model is able to produce accurate multi-structure segmentation results in only a few seconds. Different quantitative measures demonstrate an improved accuracy of our solution when compared to state-of-the-art techniques. Moreover, through a randomised survey involving expert neuroscientists, we show that subjective judgements favour our solution with respect to widely adopted atlas-based software.
\end{abstract}

\begin{keyword}
MRI \sep Brain Segmentation \sep Convolutional Neural Networks \sep Weakly Supervised Learning \sep 3D Image Analysis


\end{keyword}

\end{frontmatter}



\section{Introduction}
\label{sec:intro}

The segmentation of various brain structures from MRI scans is an essential process in several non-clinical and clinical analyses, such as the comparison at various stages of normal brain, or disease development of neurodegenerative processes, neurological diseases and psychiatric disorders.
The morphometric approach is especially helpful in pathological situations for confirming the diagnosis, defining the prognosis, and selecting the best treatment. Moreover, brain structure segmentation is an early step in functional MRI (fMRI) study pipelines, as neuroscientists need to isolate specific brain structures before analysing the spatiotemporal patterns of activity within them.
Manual segmentation, although considered to be the gold standard in terms of accuracy, is time consuming \citep{minye}. Therefore, neuroscience studies began to exploit computer vision to process data from increasingly performing MRI scanners and ease the interpretation of brain data, intrinsically characterised by a strong inter-subject variability. Different fully automated pipelines have been developed in recent years \citep{despotovicMRISegmentationHuman2015}, moving from techniques based only on image features to ones that make also use of a-priori statistical knowledge about the neuroanatomy. 
The vast majority of the available tools apply a (multi-) atlas-based segmentation strategy \citep{cabezasReviewAtlasbasedSegmentation2011}, in which the segmentation of the target volume is inferred from one or several templates built from manual annotations. In order to make this inference phase possible, a time consuming and computationally intensive \citep{reconalltime} non-rigid subject-to-atlas alignment is necessary. Due to the aforementioned high inter-subject brain variability, such registration procedures often introduce errors that yield a decrease in segmentation accuracy on brain structure or tissue boundaries \citep{kleinMindbogglingMorphometryHuman2017, lerchStudyingNeuroanatomyUsing2017a}.

In recent years, Deep Learning (DL) techniques have emerged as one of the most powerful ways to combine statistical modelling of the data with pattern recognition for decision making and classification \citep{voulodimosDeepLearningComputer2018}, and their development is impacting various medical imaging domains \citep{hamidinekoo2018deep, litjensSurveyDeepLearning2017}. Provided that they are trained on a sufficient amount of data embodying the observable variability, DL models are able to generalise well to previously unseen data. Furthermore, they can work directly with out-of-the-scanner images, removing the need for the expensive scan-to-atlas alignment phase. Numerous DL-based algorithms proposed for brain MRI segmentation match or even improve the accuracy of atlas-based segmentation tools \citep{akkusDeepLearningBrain2017a,rajchlNeuroNetFastRobust2018b,royQuickNATFullyConvolutional2018, wachingerDeepNATDeepConvolutional2018}. Due to the scarcity of training data and to hardware limitations, approaching this task using DL commonly requires the volume to be processed considering 2D \citep{royQuickNATFullyConvolutional2018} or 3D-patches \citep{fedorovEndtoendLearningBrain2016, rajchlNeuroNetFastRobust2018b, dolzHyperDenseNetHyperdenselyConnected2018b, wachingerDeepNATDeepConvolutional2018} at a time. Although this method simplifies the process from a technical point of view, it introduces significant limitations in the analysis: since each 2D or 3D patch is segmented independently from the others, these models mostly exploit local spatial information - ignoring \myquote{global} cues, such as the absolute and relative positions of different brain structures - which makes them sub-optimal. Different works have considered the potential improvements of removing said volume partitioning \citep{mcclureKnowingWhatYou2018,wachingerDeepNATDeepConvolutional2018}.
Such fully-volumetric approach has already been applied to prostate \citep{milletariVNetFullyConvolutional2016a}, heart atrium \citep{ savioliVFCNNVolumetricFully2018}, and proximal femur MRI segmentation \citep{denizSegmentationProximalFemur2018}, but not yet in the context of brain MRI segmentation - where it could prove particularly useful given the complex geometry and the variety of structures characterising the brain anatomy. 
Here, we discuss how both hardware limitations and the scarcity of hand-labelled ground truth data can be overcome. First, we tackle the former by customising and simplifying the model architecture. Second, the latter is coped with by training our model on segmentation masks obtained exploiting atlas-based techniques, in what can be considered a weakly supervised fashion. Even though this labelling is not exempt from errors, we demonstrate that the statistical reliability of atlas-based segmentation is enough to guarantee good generalisation capability of the DL models trained on such imperfect ground truth.


\section{Existing Methods for Brain MRI Segmentation and How to Advance Them}
\label{sec:prior}


\subsection{Atlas-based Methods}
\label{subsec:prior_atlas}

In the last twenty years, several atlas-based segmentation methods have been developed.
However, only few of them are completely automatic, and thus pertinent to our discussion: FreeSurfer, FSL's FAST and FMRIB, and fMRIprep.
FreeSurfer \citep{fischlFreeSurfer2012a} is an open-source software package that contains a completely automated pipeline for tissue and sub-cortical brain structure segmentation. 
FSL's FAST \citep{zhangSegmentationBrainMR2001} (FMRIB's Automated Segmentation Tool) and FIRST \citep{patenaudeBayesianModelShape2011} (FMRIB's Integrated Registration and Segmentation Tool) are part of the Oxford's open-source library of analysis tools for MRI and fMRI data. 
FAST segments different tissue types in already skull-stripped brain scans, while FIRST deals with the segmentation of sub-cortical brain structures. 
fMRIprep \citep{estebanFMRIPrepRobustPreprocessing2019} is a recently published preprocessing software for MRI scans that combines tools from widely used open-source neuroimaging packages (e.g., the above mentioned FSL and FreeSurfer). 
It implements a brain tissues segmentation pipeline, providing the user with both soft (i.e., probability maps) and hard segmentation.

These methods are widely used in neuroscience, since they produce consistent results with little human intervention.
Nevertheless, they are all atlas-based and not learning-based - hence, the only way to improve their accuracy is to manually produce new atlases.
Furthermore, since they implement a long processing pipeline together with the atlas-based labelling strategy, the segmentation operation is time consuming \citep{reconalltime}.
Limitations of these approaches, such as the lack of accuracy on various brain structure boundaries, have been documented \citep{ellingsenSegmentationLabelingVentricular2016, wengerComparingManualAutomatic2014, weierEvaluationNewApproach2012, cabezasReviewAtlasbasedSegmentation2011}.


\subsection{Deep Learning Methods}
\label{subsec:prior_dl}

The majority of the state-of-the-art methods based on deep learning exploit multi-modal MRI data \citep{cicek3DUNetLearning2016b, chenVoxResNetDeepVoxelwise2016b, dolzHyperDenseNetHyperdenselyConnected2018c, andermattMultidimensionalGatedRecurrent2016a}. 
Yet, in real-case scenarios and due to time constraints, the acquisition of different MRI sequences for anatomical analysis is rarely done: in most studies a single sequence is used - with T1\textsubscript{w} being the most popular protocol. 
Various alternatives have been proposed to obtain whole brain segmentation from T1\textsubscript{w} only. 
QuickNAT \citep{royQuickNATFullyConvolutional2018} leverages a 2D based approach to efficiently segment brain MRI, exploiting a paradigm that aggregates the predictions of three different encoder-decoder models by averaging the probability maps - each model trained to segment a single slice at a time along one of the three principal axes (longitudinal, sagittal, and coronal).
MeshNet \citep{fedorovEndtoendLearningBrain2016, mcclureKnowingWhatYou2018} is a feedforward CNN based on 3D dilated convolutions, whose structure guarantees good results while keeping the number of parameters low. 
NeuroNet \citep{rajchlNeuroNetFastRobust2018b} is an encoder-multi-decoder CNN, trained to replicate segmentation results obtained with multiple state-of-the-art neuroimaging tools. 
Finally, DeepNAT \citep{wachingerDeepNATDeepConvolutional2018} is composed of a cascade of two CNNs. 
It breaks the segmentation task into two hierarchical operations - the foreground-background separation, and the labelling of each voxel as belonging to the foreground - implemented by the first and the second network, respectively.

However, a common trait of these methods is that they do not fully exploit the 3D spatial nature of MRI data. Although QuickNAT tries to integrate spatial information by averaging the probability maps computed with respect to different views, it is slice-based. DeepNAT exploits an intrinsic parameterisation of the brain (through the Laplace-Beltrami operator) trying to introduce some spatial context, but as with MeshNet it is trained on small non-overlapping 3D-patches. Finally, NeuroNet is trained on random $128 \times 128 \times 128$ crops of the MRI.


\subsection{Aims and Contributions}
\label{subsec:aims}

\begin{figure*}[h!]
    \centering
    \includegraphics[width=0.95\textwidth]{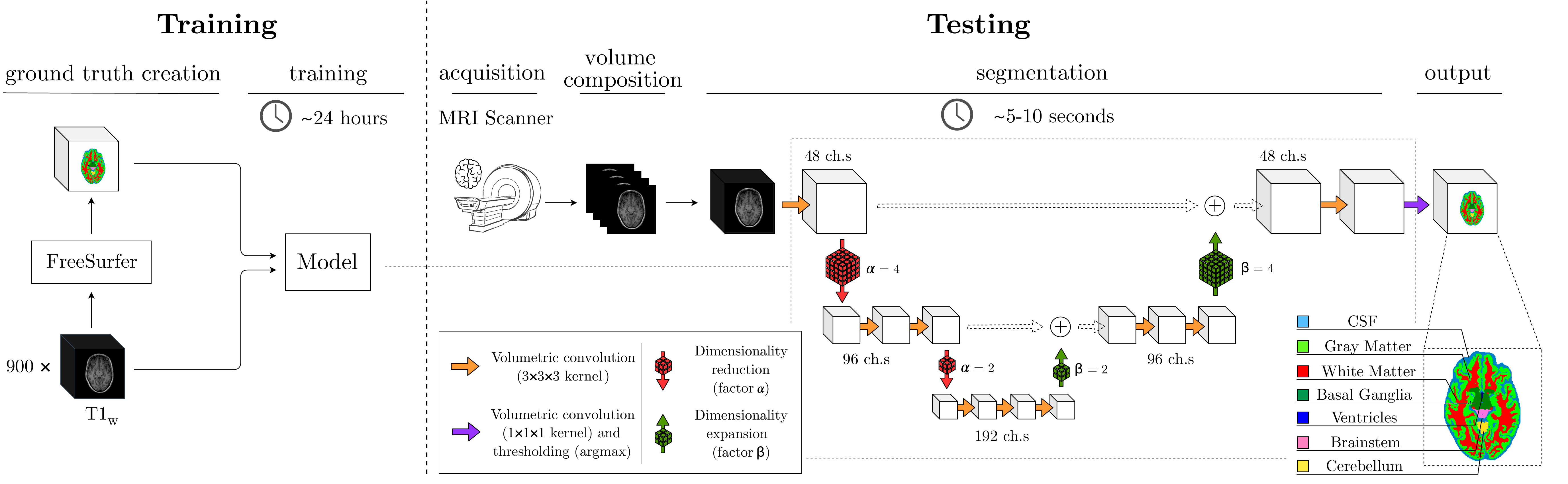}
    \caption{Overview of the proposed segmentation method. The model is trained on $900$ T1\textsubscript{w} volumes and the associated relabelled FreeSurfer segmentation, while testing is performed by feeding NIfTI data to the model.}
    \label{fig:framework}
\end{figure*}

Aiming to exploit both local and global spatial information contained in MRI data, we introduce CEREBRUM: a fast and fully-volumetric Convolutional Encoder-decodeR for weakly supervised sEgmentation of BRain strUctures from out-of-the-scanner MRI. To the best of our knowledge, CEREBRUM is the first DL model designed to tackle the brain MRI segmentation task in such a fully-volumetric fashion. This is accomplished exploiting an end-to-end encoding-decoding structure, where only convolutional blocks are used. This delivers a whole brain MRI segmentation in just $\sim$5-10 seconds on a desktop GPU\footnote{The code for training and testing will be made available on the project’s GitHub page after publication.}. The model architecture and the proposed learning framework are shown in Fig. \ref{fig:framework}.

Since in most real case scenarios, to save scanner time, only single-modal MR images are collected, we develop and test our method on a large set of data (composed by $947$ MRI scans) acquired using a T1-weighted (T1\textsubscript{w}) $1mm$ isotropic MPRAGE protocol. Neither registration nor filtering is applied to these data, so that CEREBRUM learns to segment out-of-the-scanner data. Focusing on the requirements of a real case scenario (fMRI studies), we train the model to segment the classes of interest in the MICCAI challenge \citep{mendrikMRBrainSChallengeOnline2015a} i.e., gray matter (GM), white matter (WM), cerebrospinal fluid (CSF), ventricles, cerebellum, brainstem, and basal ganglia. 
Since manually annotating such a large body of data would require a prohibitive amount of human hours, we train our model on automatic segmentations obtained by FreeSurfer \citep{fischlFreeSurfer2012a} - relabelled to obtain the aforementioned set of seven classes.

We compare the proposed method with other CNN-based solutions: the well-known 2D-patch-based U-Net \citep{ronnebergerUNetConvolutionalNetworks2015a}, its 3D variant \citep{cicek3DUNetLearning2016b}, and the state-of-the-art architecture QuickNAT \citep{royQuickNATFullyConvolutional2018} - which leverages the aggregation of three slightly modified U-Net architectures (trained on coronal, sagittal, and axial MRI slices, respectively).
To ensure a fair comparison, we train these models by conducting an extensive hyperparameter selection process. 
Results are quantitatively evaluated exploiting the same metrics used in the MICCAI MR Brain Segmentation challenge, i.e., the Dice 
Similarity Coefficient, the $95^{th}$ Hausdorff Distance, and the Volumetric Similarity Coefficient \citep{metrics2015}, utilising FreeSurfer as GT reference. In addition, to assess the generalisation capability of the proposed model, we compare the obtained results against the FreeSurfer segmentation we used for training. To do so, we design a survey\footnote{The code for the qualitative test will be made available on the project’s GitHub page after publication.} in which five expert neuroscientists (with more than five years of experience in MRI analysis) are asked to choose the most accurate segmentation between the two aforementioned ones. This qualitative test covers different areas of interest in neuroimaging studies, i.e., the early visual cortex (EVC), the high-level visual areas (HVC), the motor cortex (MCX), the cerebellum (CER), the hippocampus (HIP), the early auditory cortex (EAC), the brainstem (BST) and the basal ganglia (BGA).


\section{Data}
\label{sec:data}
To speed up research and promote reproducibility, numerous large-scale neuroimaging experiments make the collected data available to all researchers \citep{OASIS07, vanessenWUMinnHumanConnectome2013, abcd2019, ukbb2016, adhd2017}. However, none of these studies provide a manually annotated ground-truth (GT), as carrying out the operation on such large databases would prove exceptionally time-consuming.

For this reason, most of the studies investigating the application of Deep Learning architectures for brain MRI segmentation make use of automatically produced GT for training purposes \citep{royQuickNATFullyConvolutional2018, mcclureKnowingWhatYou2018, fedorovEndtoendLearningBrain2016, rajchlNeuroNetFastRobust2018b} - with some of them reporting the latter can be exploited to train models that perform the same \citep{rajchlNeuroNetFastRobust2018b}, or even better \citep{royQuickNATFullyConvolutional2018}, than the automated pipeline itself. Motivated by this rationale, to train and test the proposed model we exploit a large collection of out-of-the-scanner MR images and the results of the FreeSurfer \citep{fischlFreeSurfer2012a} cortical reconstruction process \texttt{recon-all} as reference GT. As anticipated in Section \ref{sec:intro}, we relabel this result preserving seven among the most important classes of interest in most of fMRI studies (see Section \ref{subsec:aims} and Fig. \ref{fig:framework}).

\begin{figure*}
    \centering
    \subfigure[Subject 1]{
    	\includegraphics[width=0.5\textwidth, trim={0cm, 0cm, 1cm, 0cm}, clip]{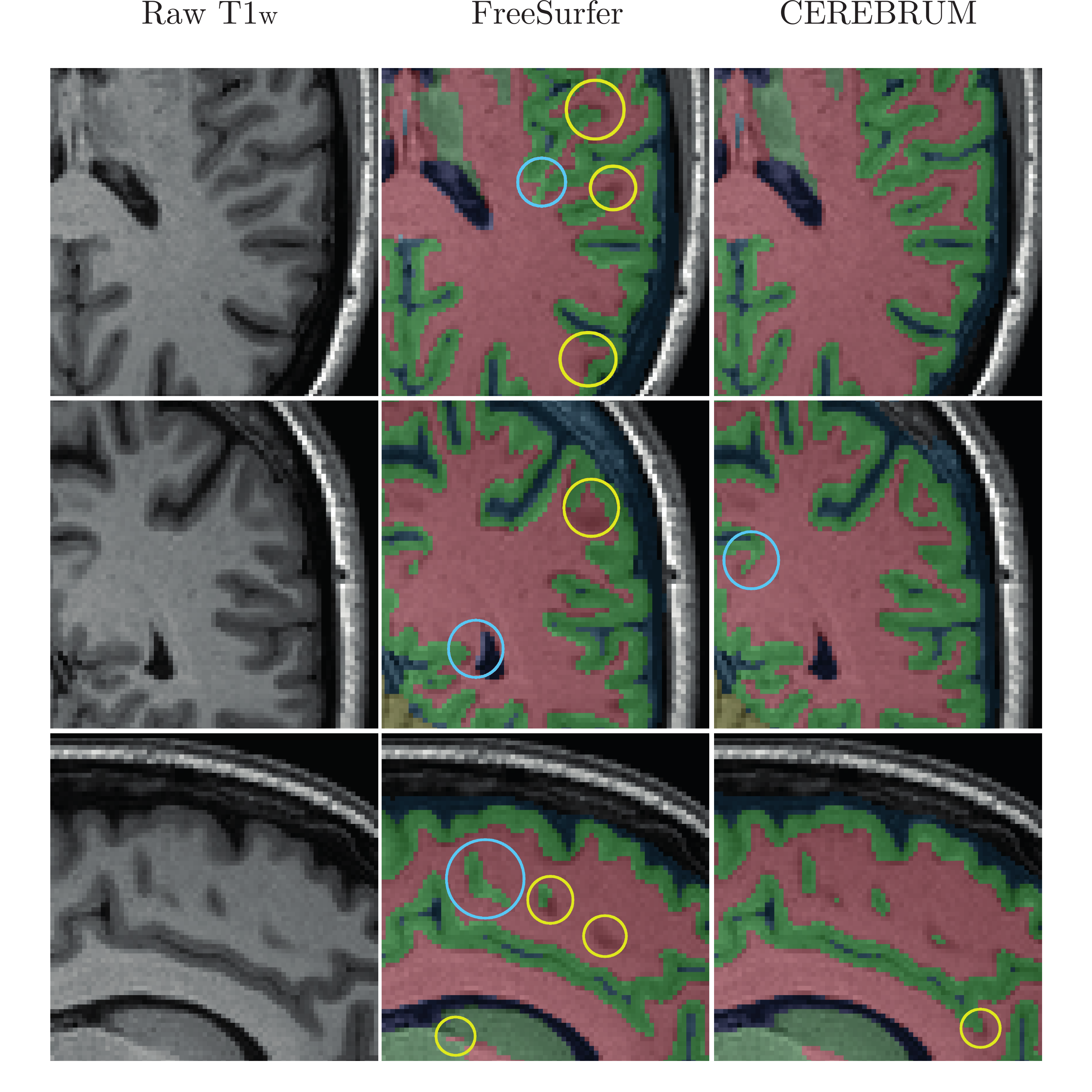}
    	\label{fig:ccni_slices_a}}%
    \subfigure[Subject 4]{
    	\includegraphics[width=0.5\textwidth, trim={1cm, 0cm, 0cm, 0cm}, clip]{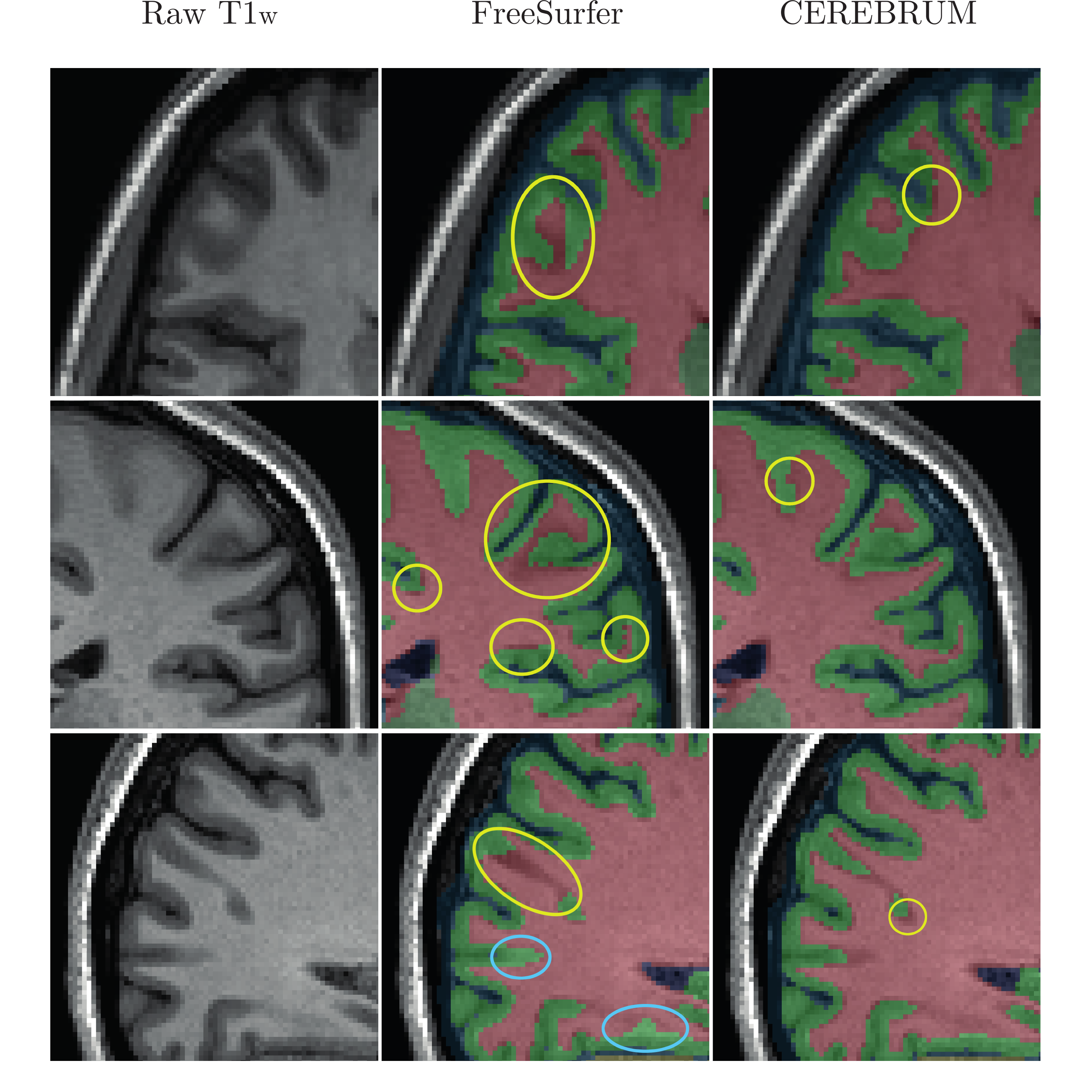}
    	\label{fig:ccni_slices_b}}
    \caption{Out-of-the-scanner (contrast enhanced) T1\textsubscript{w} scan (left), FreeSurfer segmentation (middle), and the result produced by our model (right). Fig. (a) depicts slices of test Subject 1, while (b) slices of test Subject 4 (sagittal, coronal, and longitudinal view, respectively). Cases of white matter over-segmentation are highlighted by yellow circles, while cases of white matter under-segmentation are highlighted by turquoise circles (best viewed in electronic format).}
        \label{fig:ccni_slices}
\end{figure*}
The database, collected from the Centre for Cognitive Neuroimaging (the University of Glasgow) in more than $10$ years of machine activity, consists of $947$ MR images - $900$ of which are used for training, $11$ for validation, and $36$ for testing. All the volumes are out-of-the-scanner, i.e, obtained directly from a set of DICOM images using \texttt{dcm2niix} \citep{dcm2niix}, whose auto-crop option is exploited to make sizes consistent across all the dataset (i.e., $192\times256\times170$ for sagittal, coronal, and longitudinal axis, respectively) without pre-processing the data. Given the number of available scans for training, and since no atlas-based spatial normalization is performed, the actual variability in shape, rotation, position and anatomical size is such that no data augmentation is needed to avoid the risk of overfitting.
The first two columns of Fig. \ref{fig:ccni_slices_a} and \ref{fig:ccni_slices_b} show detailed views from some selected slices of the out-of-the-scanner T1\textsubscript{w} and the corresponding relabelled FreeSurfer segmentation, respectively. The main characteristics of the dataset are summarised in Table \ref{tab:dataset}.
As the data have been collected under different ethics applications, we are not able to make the whole database publicly available. However, $7$ out of $36$ volumes used for testing are collected under the approval of the local ethics committee of the College of Science \& Engineering (ethics \#$300170016$) and shared online after anonymisation\footnote{Data will be shared on openneuro.org after publication.}, for comparison and research purposes, along with the segmentation masks resulting from CEREBRUM and FreeSurfer (See Figure \ref{fig:ccni_slices} and Section \ref{subsec:qualitative}).

\begin{table}[h]
\caption{Datasets details. MR Images acquired at the Centre for Cognitive Neuroimaging (University of Glasgow, UK)} 
\centering 
\begin{tabular}{C{4.5cm}  C{3.5cm}} 

\toprule
	\textbf{Parameter} & \textbf{Value} \\
\midrule
\midrule
	Sequence used & T1\textsubscript{w} MPRAGE \\	\midrule
	Field strenght & 3 Tesla \\	\midrule
	Voxel size & 1mm-isotropic\\	\midrule
	Volume sizes (original) & $192\times256\times256$ \\	\midrule
	Volume sizes (preprocessed) & $192\times256\times170^\dagger$ \\		\midrule
	Training & $900$ volumes \\ 	\midrule
	Validation & $11$ volumes \\ 	\midrule
	Testing & $36$ volumes\textsuperscript{*} \\   \bottomrule	
\multicolumn{2}{l}{\textsuperscript{*} 7 of which are publicly available.}\\
\multicolumn{2}{l}{$^\dagger$ out-of-the-scanner data, neck cropping only.}
	
\end{tabular}
\label{tab:dataset} 
\end{table}


\section{Proposed model}
\label{sec:models}

To make the complexity of managing our $192\times256\times170$ voxels data tractable, we carefully optimise the model architecture so as to implicitly deal with GPU memory constraints. Furthermore we exploit, for training purposes, a machine equipped with $4$ GeForce\textsuperscript{\textregistered} GTX $1080$ Ti.

Inspired by \cite{ronnebergerUNetConvolutionalNetworks2015a} and \cite{cicek3DUNetLearning2016b}, we propose a deep encoder-decoder model with six 3D convolutional blocks, which are arranged in increasing number on three layers.
Since a whole volume is considered as an input, the feature maps extracted by such convolutional blocks are not limited to patches but span across the entire volume. As each block captures the content of the whole brain MRI, this enables the learning of both local and global spatial features by leveraging the spatial context which is propagated to each subsequent block.
Furthermore, in order to better exploit the fine details found in 3T brain MRI data, kernels of size $3\times3\times3$ are used as feature extractors.
Instead of max-pooling, convolutions with stride are used as a dimensionality reduction method, thus allowing the network to learn the optimal down sampling strategy starting from the extracted features. Exploiting such operations, and to force the learning of more abstract (spatial) features, a factor $1:64$ dimensionality reduction is implemented after the first layer.
Finally, skip connections are used along with tensorial sum (instead of concatenation) to improve the quality of the segmented volume while significantly limiting the number of parameters \cite{quanFusionNetDeepFully} to $\sim 5M$, far less with respect to state-of-the-art models which are structured in a similar fashion.

We train the model by optimising the categorical cross-entropy function. Convergence is achieved after roughly $24$ hours of training ($40$ epochs), using Adam \cite{adam2014} with a learning rate of $42\cdot 10^{-5}$, $\beta_1 = 0.9$ and $\beta_2 = 0.999$.
Furthermore, we set the batch size to $1$ and thus do not implement batch normalisation \cite{IS15}.


\section{Results}
\label{sec:results}

The results we present in this section aim to confirm the hypothesis that avoiding the partitioning of MRI data enables the model to better learn global spatial features useful for segmentation. At first, in Section~\ref{subsec:quantitative}, we provide numerical comparison with other state-of-the-art CNN architectures (U-Net \cite{ronnebergerUNetConvolutionalNetworks2015a}, 3D U-Net \cite{cicek3DUNetLearning2016b}, QuickNAT \cite{royQuickNATFullyConvolutional2018}). Then, in Section~\ref{subsec:qualitative}, we conduct a survey involving expert neuroscientists to subjectively assess the CEREBRUM segmentation accuracy. Finally, we further verify the validity of our assumptions by inspecting the soft-segmentation maps produced by the models in Section~\ref{subsec:prob_maps}, and we demonstrate the suitability of our dataset by analysing the impact of the training set size on CEREBRUM performance in Section~\ref{subsec:result_costum}.


\subsection{Numerical Comparison}
\label{subsec:quantitative}

We numerically assess the performance of the models, using FreeSurfer segmentation as a reference, exploiting the metrics utilised in the MICCAI MRBrainS18 challenge (among the most employed in the literature \citep{metrics2015}). Dice (similarity) Coefficient (DC) is a measure of overlap, and a common metric in segmentation tasks. The Hausdorff Distance, a dissimilarity measure, is useful to gain some insight on contours segmentation. Since HD is generally sensitive to outliers, a modified version (95\textsuperscript{th} percentile, HD95) is generally used when dealing with medical image segmentation evaluation \citep{hausdorff1992}. Finally, the Volumetric Similarity (VS), as the name suggests, evaluates the similarity between two volumes.

\begin{figure*}

    \centering
    \includegraphics[width=0.86\textwidth]{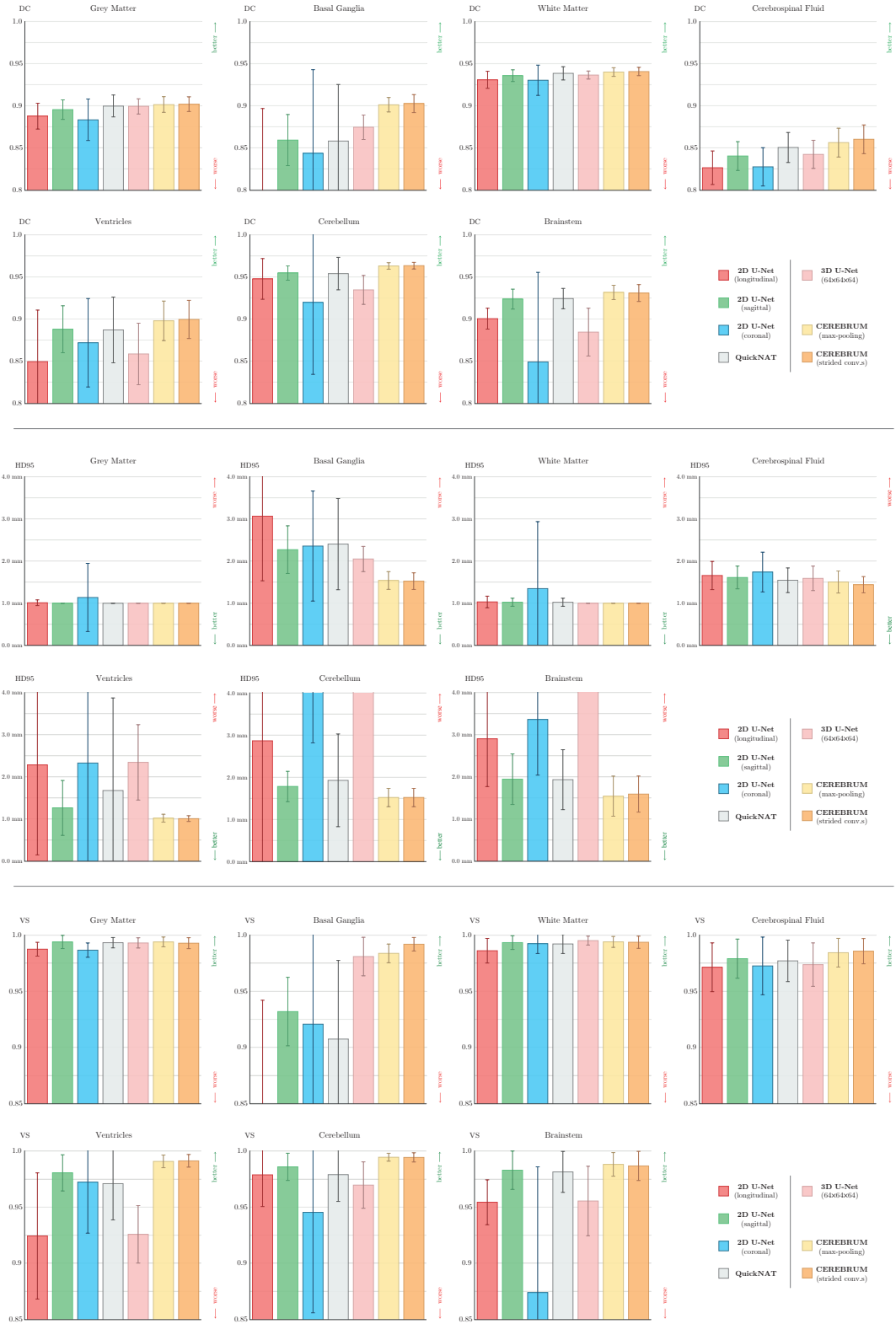}
    
    \caption{Dice Coefficient, 95\textsuperscript{th} percentile Hausdorff Distance, and Volumetric Similarity computed using FreeSurfer relabelled segmentation as a reference. The 2D-patch-based (red, green, blue and grey for longitudinal, sagittal, coronal, and view-aggregation, respectively), the 3D-patch-based (pink), and our model (yellow for max-pooling and orange for strided convolutions) are compared. The height of the bar indicates the mean across all the test subjects, while the error bar represents the standard deviation (best viewed in electronic format).}
    \label{fig:numerical_results}
\end{figure*}

CEREBRUM is compared against state-of-the-art encoder-decoder architectures: the well-known-2D-patch based U-Net \cite{ronnebergerUNetConvolutionalNetworks2015a} (trained on the three principal views, i.e., longitudinal, sagittal and coronal), the 3D-patch based U-Net 3D \cite{cicek3DUNetLearning2016b} (with 3D patches sized $ 64 \times 64 \times 64$ \cite{cicek3DUNetLearning2016b, fedorovEndtoendLearningBrain2016,PKL17}), and the QuickNAT \cite{royQuickNATFullyConvolutional2018} architecture (which implements view-aggregation starting from 2D-patch based models).
We train all the models minimising the same loss for $50$ epochs, using the same number of volumes, and similar learning rates (with changes in those regards made to ensure the best possible validation score).
Figure \ref{fig:numerical_results} shows class-wise results (DC, HD95 and VS) depicting the average score (computed across all the $36$ test volumes) and the standard deviation.
We compare 2D-patch-based (longitudinal, sagittal, coronal), QuickNAT, 3D-patch-based, and CEREBRUM (both a max pooling and strided convolutions version).
Overall, the latter outperforms all the other CNN-based solutions on every class, despite having far less parameters: when its average score (computed across all the subjects) is comparable with that of other methods (e.g., view-aggregation, GM), it has a smaller variability (suggesting higher reliability).


\subsection{Experts' Qualitative Evaluation}
\label{subsec:qualitative}

The quantitative assessment presented in Section \ref{subsec:quantitative}, though informative, cannot be considered exhaustive. Indeed, using FreeSurfer as a reference for such evaluation makes the latter a ranking on a relative scale - and if this highlights the value of the fully-volumetric approach, it does not make a direct comparison with the atlas-based method possible. Thus, we need to confirm more systematically what can be inferred, for instance, from Figure \ref{fig:ccni_slices} - where far superior qualitative performance of CEREBRUM are clear compared to FreeSurfer, as the former produces more accurate segmentation masks, with far less holes and bridges. This somehow surprising generalisation capability of CEREBRUM over its training reference, if confirmed, would prove the desired \myquote{strengthening} effect yielded by the adoption of a weakly supervised learning approach.
Moreover, quantitative assessments are often criticised by human experts, such as physicians and neuroscientists, for they do not take into account the severity of each segmentation error \cite{metrics2015}, which is of critical importance in professional usage scenarios. 

\begin{figure}[H]
\centering
	\includegraphics[width=0.75\textwidth, trim={0, 2cm, 0cm, 1cm}, clip]{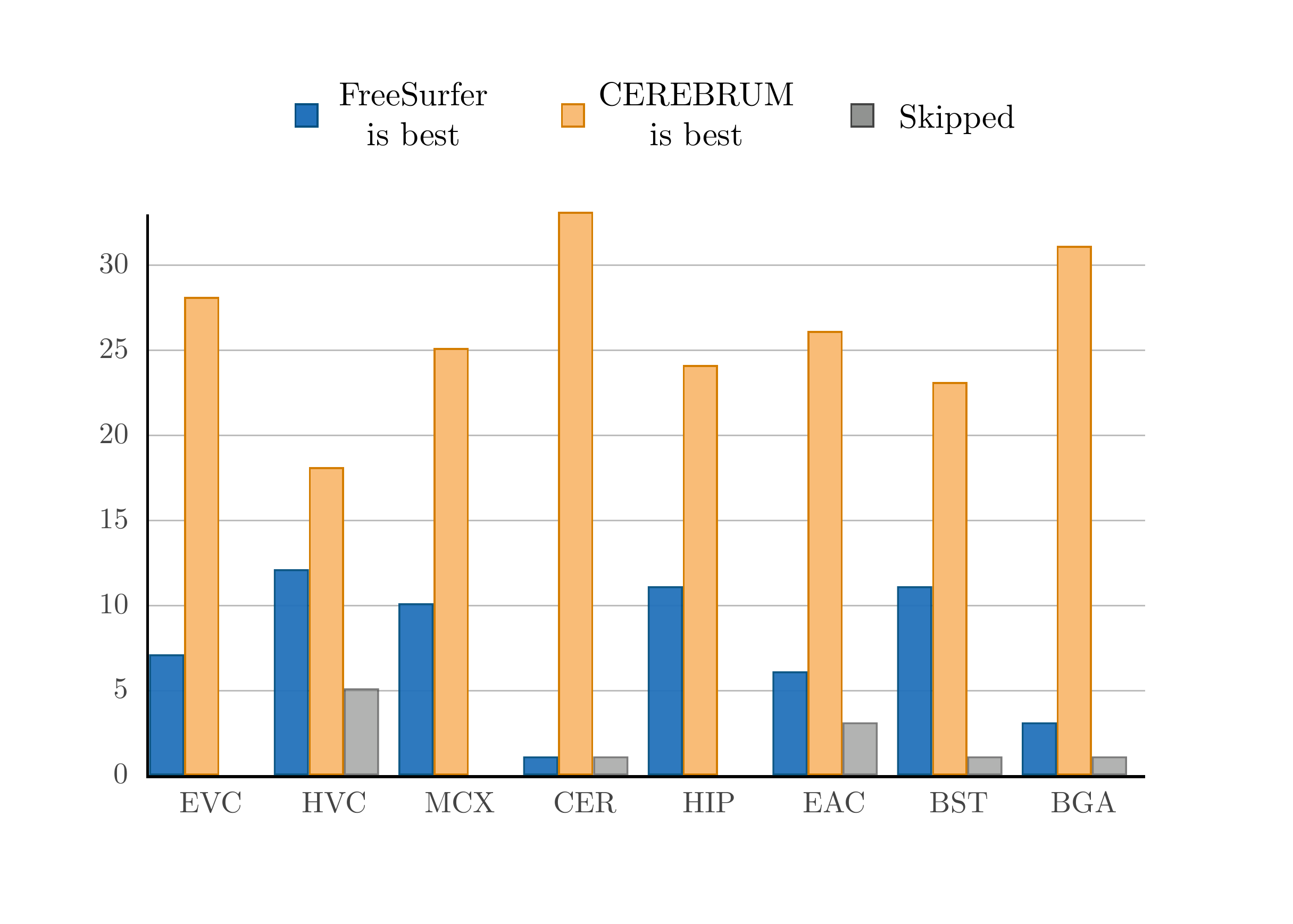}
	
	\medskip
	
    \caption{Outcome of the segmentation accuracy assessment test, conducted by expert neuroscientists, for the following areas: early visual cortex (EVC), the high-level visual areas (HVC), the motor cortex (MCX), the cerebellum (CER), the hippocampus (HIP), the early auditory cortex (EAC), the brainstem (BST) and the basal ganglia (BGA). The bars represent the number of preferences expressed by the experts: CEREBRUM (in orange), FreeSurfer (in blue), or none of the two (in grey).}
    \label{fig:experts_evaluation}
\end{figure}

For the aforementioned reasons, we design and implement a systematic subjective assessment by means of a PsychoPy \cite{psychopy2007} test in which five expert neuroscientists (with more than five years of expertise in MRI analysis) are asked to choose the most accurate segmentation between the one produced by CEREBRUM and the (relabelled) FreeSurfer one. The participants are presented with a coronal, sagittal, or axial slice selected from a test volume, and are allowed both to navigate between four neighbouring slices (two following and two preceding the displayed one) and to change the opacity of the segmentation mask (from $0\%$ to $100\%$) to better evaluate the latter with respect to the anatomical data. This process is repeated seven times - one for each test subject - per each of the eight brain areas of interest, i.e., early visual cortex (EVC), the high-level visual areas (HVC), the motor cortex (MCX), the cerebellum (CER), the hippocampus (HIP), the early auditory cortex (EAC), the brainstem (BST) and the basal ganglia (BGA). The choice of the slices to present and the order in which the latter are arranged is randomised. Furthermore, the neuroscientists are allowed to skip as many slices as they want if they are unsure about the choice: such cases are reported separately.
From the results shown in Figure \ref{fig:experts_evaluation} it emerges that, according to expert neuroscientists, CEREBRUM qualitatively outperforms FreeSurfer. This proves the model superior generalisation capability and provides evidence to support the adopted weakly supervised approach. Moveover, such results hint at the possibility to have atlas-based methods and deep learning ones operating together in a synergistic way.

\subsection{Probability Maps}
\label{subsec:prob_maps}

To further investigate the hypothesis that a fully-volumetric approach is advantageous with respect to other patch-based models, we also conduct a qualitative assessment on the predicted probability maps (i.e., soft segmentation).
Such evaluation could clearly reveal the ability of the model to make use of spatial cues: for instance, a well-learned model which exploits learned spatial features should predict the presence of cerebellum voxels only in the back of the brain, where the structure is normally located.

\begin{figure*}[h!]
    \centering
    \subfigure[Cerebellum (sagittal view)]{
    	\includegraphics[width=0.5\textwidth, trim={0cm, 0cm, 0cm, 0cm}, clip]{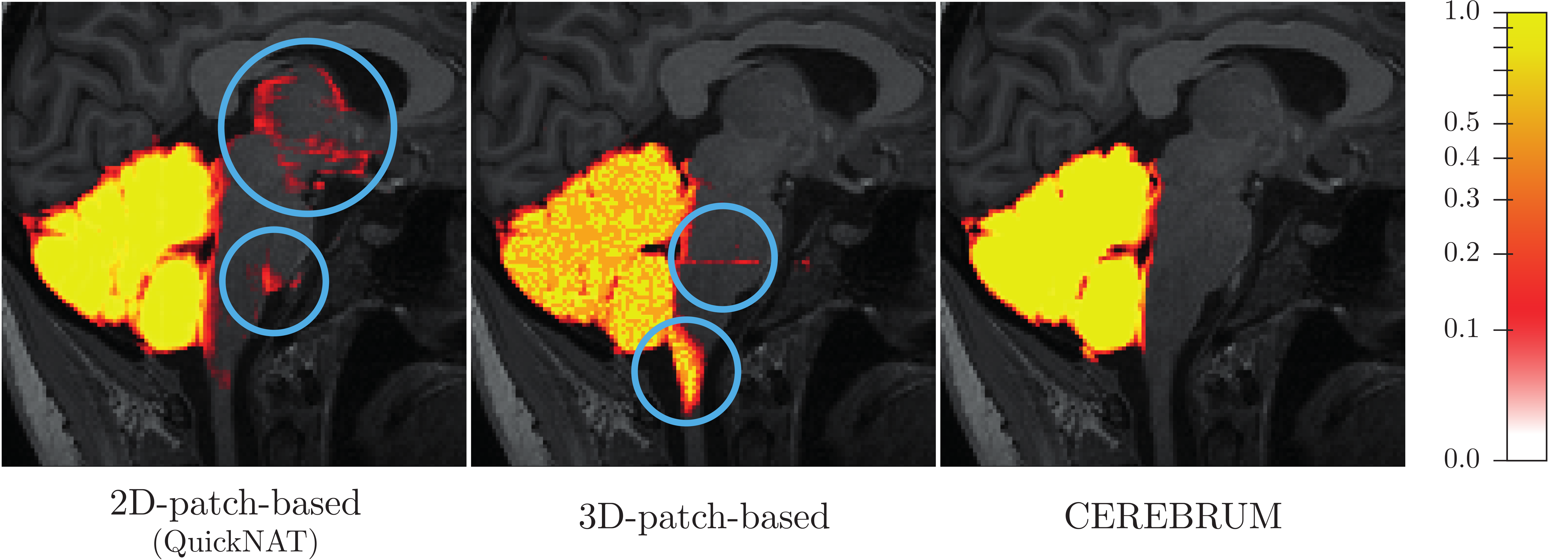}
    	\label{fig:soft-seg-a}}%
    \subfigure[Basal ganglia (longitudinal view)]{
    	\includegraphics[width=0.5\textwidth, trim={0cm, 0cm, 0cm, 0cm}, clip]{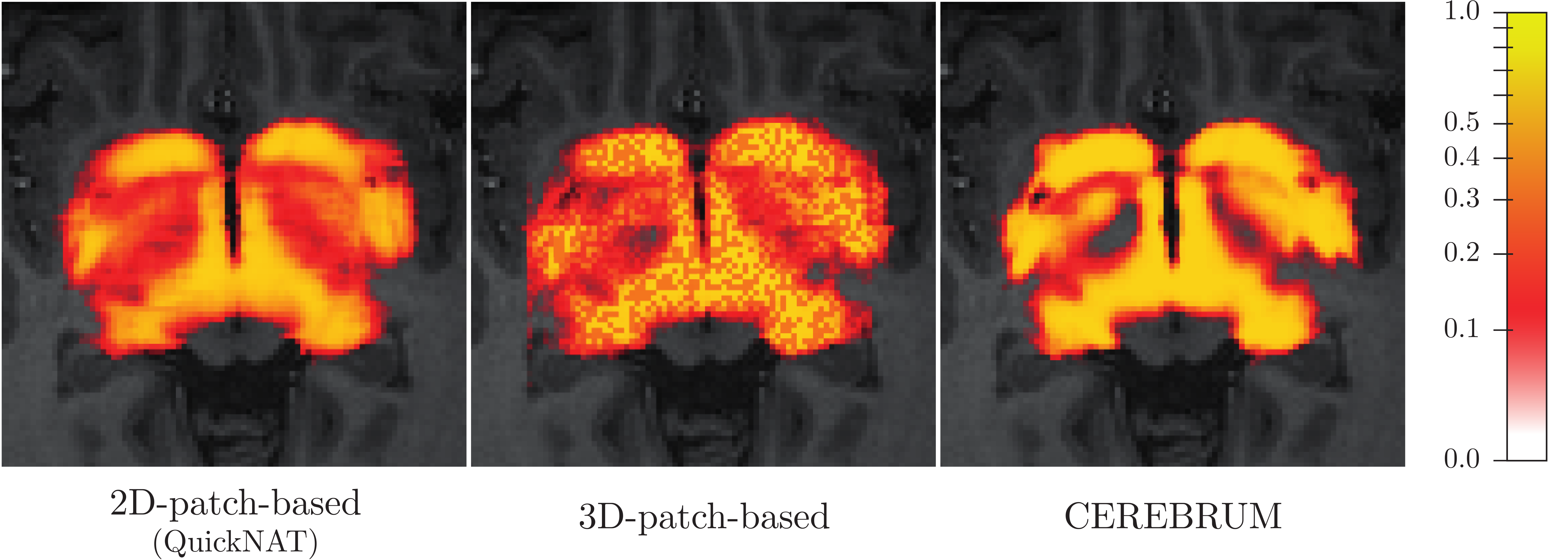}
    	\label{fig:soft-seg-b}}
    \caption{Soft segmentation maps of subject 1 cerebellum (a) and the basal ganglia (b) produced by the best 2D-patch-based model (QuickNAT), the 3D-patch-based model (3D U-Net), and CEREBRUM (ours). The proposed approach produces results that are spatially more coherent, and lack of false positives (highlighted in light blue; best viewed in electronic format).}
        \label{fig:soft-seg}
\end{figure*}

Figure \ref{fig:soft-seg-a} and \ref{fig:soft-seg-b} show two selected slices of the soft segmentation (percent probability, displayed in logarithmic scale) resulting from the best 2D-patch-based method (i.e., QuickNAT), the 3D-patch-based method, and CEREBRUM - for the cerebellum and basal ganglia classes, respectively (superimposed to the corresponding T1\textsubscript{w} slice). Other classes are omitted for clarity.

The probability maps produced by the 2D and 3D-patch based methods are characterised by the presence of voxels associated with significant probability of belonging to the structure of interest ($p > 0.2$) despite their distance from the latter. This can lead to misclassification errors in the hard segmentation (after the thresholding). In particular, higher uncertainty and spurious activations due to views averaging can be seen in the soft segmentation maps produced by QuickNAT - while blocking artefacts on the patch borders are visible in the case of the 3D-U-Net, even when the latter is trained using overlapping 3D-patches whose predictions are then averaged.
The soft segmentation produced by CEREBRUM, on the contrary, is more coherent and closer to the reference in both cases and does not present the aforementioned errors. This hints to the superior ability of the proposed model in learning both global and local spatial features.

\subsection{Number of Training Samples}
\label{subsec:result_costum}

\begin{figure}[H]
    \centering
    \includegraphics[width=0.5\textwidth]{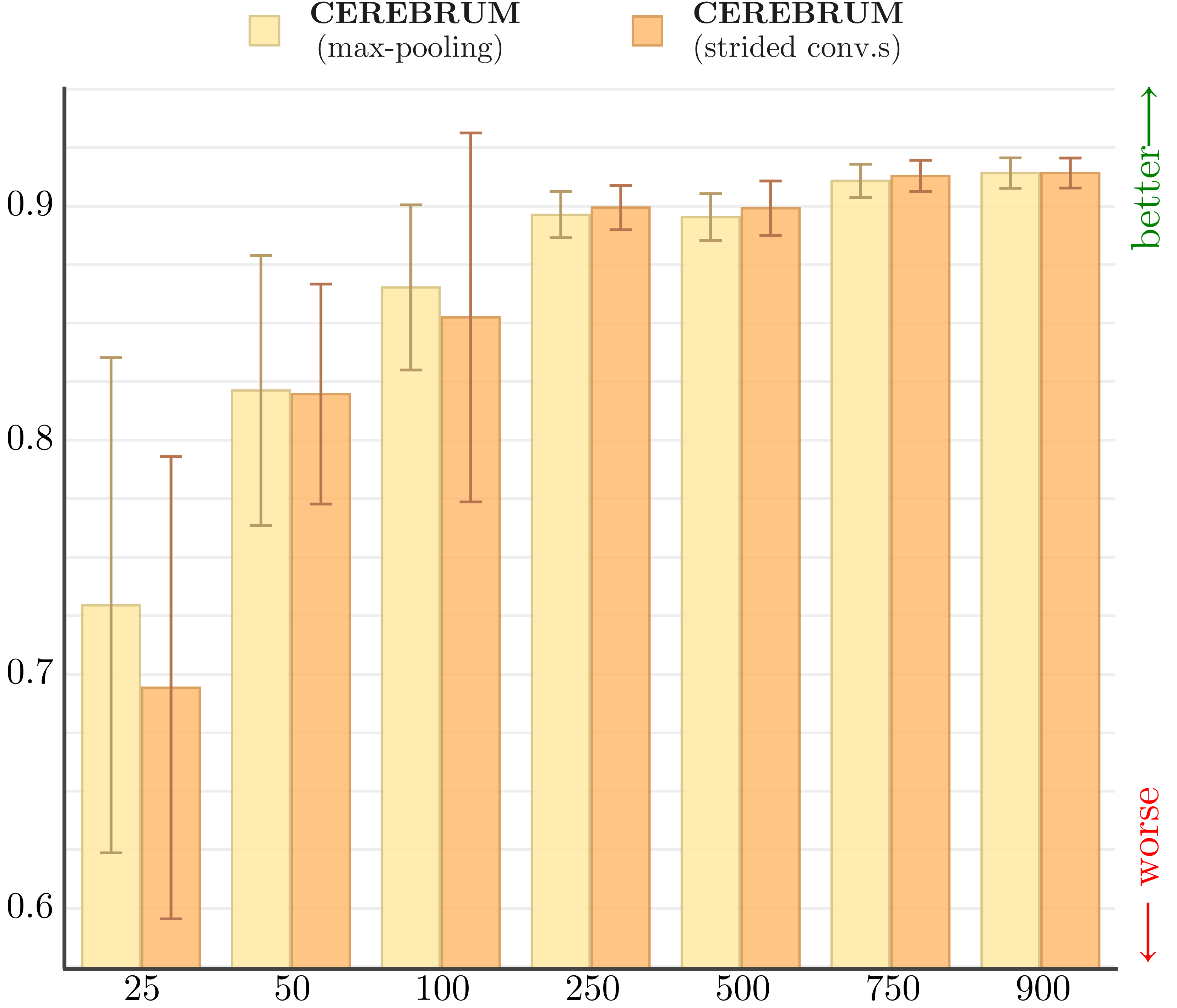}
    
    \medskip
    \caption{Impact of the training set size on the performance - Dice Coefficient averaged across all the seven classes. Results are computed on the whole test set ($36$ volumes).}
    \label{fig:n_samples}

\end{figure}

One of the possible limitations of approaching the brain MRI segmentation task in a fully-volumetric fashion could be the scarcity of training data - for in such a case each volume does not yield many training samples, as for 2D and 3D-patch-based solutions, but a single one. To investigate this possible drawback, we evaluate the performance of CEREBRUM when trained on smaller sub-sets of our database.
In particular, we train the proposed model by randomly extracting $25, 50, 100, 250, 500, 700, 900$ samples from the training set. To evaluate the performance of the model in the first two cases (i.e., $25$ and $50$ MRI scans), we repeat the training $5$ times (on randomly extracted yet non-overlapping subsets of the database) and average the results.
Furthermore, we evaluate the impact on the performance yielded by the introduction of strided convolutions (i.e., more learnable parameters) when the training set size is limited by training a variation of CEREBRUM where max-pooling is used as a dimensionality-reduction strategy.
Figure \ref{fig:n_samples} shows that the performance variation significantly deteriorates as the training set size falls below $250$ samples, while substantial stability is reached over $750$ samples. This is a confirmation that our $900$ samples training set is properly sized to the target without there being any urge for data augmentation.

\section{Conclusion}
\label{sec:discussion}

In this work we presented CEREBRUM, a CNN-based deep model that approaches the brain MRI segmentation problem in a fully-volumetric fashion. The proposed architecture is a carefully (architecturally) optimised encoder-decoder that, starting from a T1\textsubscript{w} MRI volume, produces a result in only few seconds on a desktop GPU.
We evaluated the proposed model performance, comparing it to state-of-the-art 2D and 3D-patch-based models with similar structure, exploiting the Dice Coefficient, the 95\textsuperscript{th} percentile Hausdorff Distance, and the Volumetric Similarity, assessing CEREBRUM superior performance. Furthermore, we conducted a survey of expert neuroscientists to obtain their judgements about the accuracy of the resulting segmentation, comparing the latter with the result of FreeSurfer cortical reconstruction process. According to the participants to such experiment, CEREBRUM achieves better segmentation than FreeSurfer. 
To our knowledge, this is the first time a DL-based fully-volumetric approach for brain MRI segmentation is deployed. The results we obtained prove the potential of this approach, as CEREBRUM outperforms 2D and 3D-patch-based encoder-decoder models using far less parameters. Removing the partitioning of the volume, as hypothesised, allows the model to learn both local and spatial features. 
Furthermore, we are also the first conducting a qualitative assessment test consulting expert neuroscientists: this is fundamental, as commonly used metrics often fail to capture the information experts need to rely on DL methods and exploit the latter for research.  

\section*{Acknowledgements}

This project has received funding from the European Union’s Horizon 2020 Programme for Research and Innovation under the Specific Grant Agreement No. 785907 (Human Brain Project SGA2) awarded to LM.





\bibliographystyle{model1-num-names}
\bibliography{bibliography.bib}







\end{document}